\documentclass[aps,twocolumn,showpacs,superscriptaddress]{revtex4}
\usepackage{amsmath}
\usepackage{graphicx}
\usepackage{esint}
\usepackage{epstopdf}%This line makes .eps figures into .pdf - please comment out if not required.
\graphicspath{{pict/}{./}}
\usepackage{bm}%
\newcounter{Fig}

\newcommand{\be}{\begin{equation}}
\newcommand{\ee}{\end{equation}}

\begin{document}

%\title{Q-factor enhancement for all-dielectric nanoresonators through relaxed total internal reflection}
\title{Superscattering pattern shaping for radially anisotropic nanowires}
\author{Wei Liu}
\email{wei.liu.pku@gmail.com}
%\footnote{wei.liu.pku@gmail.com}
\affiliation{College of Optoelectronic Science and Engineering, National University of Defense
Technology, Changsha, Hunan 410073, P. R. China}
%\author{TBD}
%\affiliation{Nonlinear Physics Centre,  Research
%School of Physics and Engineering, Australian National University,
%Canberra, ACT 0200, Australia}
%\author{Andrey E. Miroshnichenko}
%\affiliation{Nonlinear Physics Centre,  Research
%School of Physics and Engineering, Australian National University,
%Canberra, ACT 0200, Australia}
%\author{Yuri S. Kivshar}
%\affiliation{Nonlinear Physics Centre,  Research
%School of Physics and Engineering, Australian National University,
%Canberra, ACT 0200, Australia}
%\affiliation{Department of Nanophotonics and Metamaterials, ITMO University, St. Petersburg 197101, Russia}

\begin{abstract}
We achieve efficient shaping of superscattering by radially anisotropic nanowires relying on resonant multipolar interferences. It is shown that the radial anisotropy of refractive index can be employed to resonantly overlap electric and magnetic multipoles of various orders, and as a result effective superscattering with different engineered angular patterns can be obtained. We further demonstrate that such superscattering shaping relying on unusual radial anisotropy parameters can be directly realised with isotropic multi-layered nanowires, which may shed new light to many fundamental researches and various applications related to scattering particles.
\end{abstract}
\maketitle

\section{Introduction}

Stimulated by the recent demonstrations of optically-induced magnetic responses in various nanostructures incorporating high refractive index materials~\cite{Liu2014_CPB,jahani_alldielectric_2016,KUZNETSOV_Science_optically_2016}, the principle of multipolar interferences has been widely applied in many particle scattering related systems and  metasurfaces, provoking lots of applications and fundamental mechanism investigations in both the linear and nonlinear regimes~\cite{SMIRNOVA_Optica_multipolar_2016,LIU_ArXivPrepr.ArXiv160901099_multipolar_2016}. Among all the multipolar interference effects, an outstanding and attractive example is the efficient shaping of the superscattering that is beyond the single channel scattering limit~\cite{Ruan2010_PRL,Liu2014_arXiv_Geometric,LI_IEEEJ.Sel.Top.QuantumElectron._design_2017}. Such a manipulation relies on the resonant overlapping and interferences of electric and magnetic resonances of various orders ~\cite{Liu2014_CPB,jahani_alldielectric_2016,KUZNETSOV_Science_optically_2016,SMIRNOVA_Optica_multipolar_2016,LIU_ArXivPrepr.ArXiv160901099_multipolar_2016}, and can play a vital role in many applications that require simultaneous strong scattering and designed angular scattering distributions.

Conventional methods to resonantly overlap different multipoles supported by an individual particle reply on: (i) a flat dispersion band  or multiple dispersion bands to overlap electric resonances of different orders ~\cite{Ruan2010_PRL,Liu2014_arXiv_Geometric,LI_IEEEJ.Sel.Top.QuantumElectron._design_2017} and (ii) metal-dielectric hybrid nanostructures~\cite{Liu2012_ACSNANO,mirzaei2013cloaking,Liu2013_OL2621,Liu2014_ultradirectional} or homogeneous dielectric particles of irregular shapes that are neither spherical nor cylindrical~\cite{Staude2013_acsnano,LUKYANCHUK_ACSPhotonics_optimum_2015} to overlap electric and magnetic resonances. Recently it is shown that for the fundamental homogeneous spherical particles, effective radial index anisotropy can be applied to significantly tune the positions of the electric resonances of various orders~\cite{Liu2015_OE_Ultra,LIU_Sci.Rep._unidirectional_2016}. This enables flexible overlapping of multipoles of different natures (electric or magnetic) and/or orders, which results in multipolar interference induced superscattering of different angular scattering distributions, including the ultra-directional forward superscattering~\cite{Liu2015_OE_Ultra,LIU_Sci.Rep._unidirectional_2016}).

In this work, we extend our investigations of radial anisotropy induced superscattering pattern shaping from three dimensional (3D) spherical particles~\cite{Liu2015_OE_Ultra,LIU_Sci.Rep._unidirectional_2016} to two dimensional (2D) cylindrical ones. Such an extension is by no means trivial, considering that compared to its 3D counterparts, the scattering of 2D nanowires shows polarization dependence and more importantly, the number of degenerate scattering channels are quite different from those of 3D spherical particles for each resonance~\cite{Liu2012_ACSNANO,Liu2013_OL2621,Kerker1969_book}. Here in this work we have achieved simultaneously, within radially anisotropic nanowires, flexible tuning of resonance positions and superscattering with different engineered scattering patterns induced by the interferences of multipoles resonantly overlapped.  It is further shown that such efficient superscattering pattern shaping replying on naturally inaccessible anisotropy parameters can be simply realised within isotropic multi-layered nanowires.  Such an approach based on refractive index anisotropy renders an extra dimension of freedom for resonance tuning and scattering pattern shaping, which might play a significant role in various investigations into light-matter interactions and plenty of applications associated with particle scattering, such as nanoantennas, optical sensing and detections, scattering particle assisted passive radiation cooling and photovoltaic devices~~\cite{Novotny2012_book,ZHAI_Science_scalablemanufactured_2017-1}.

\begin{figure}
\centerline{\includegraphics[width=9cm]{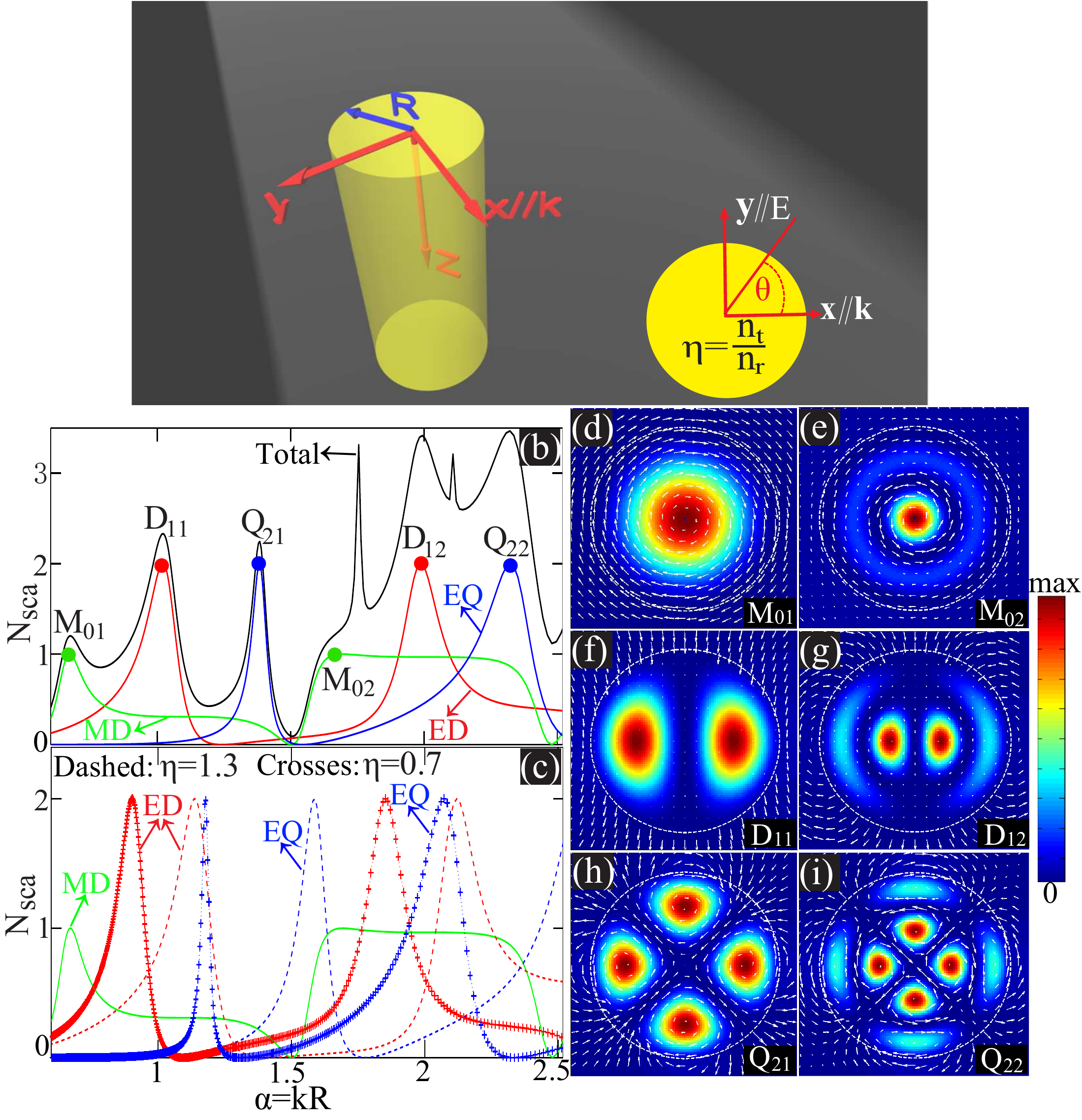}} \caption {(a) Schematic illustration of a normally incident plane wave scattererd by a radially anisotropic nanowire of radius $R$. The incident wave is TM polarized, with magnetic field along $z$ direction and wave-vector $\textbf{k}$  along $x$ direction. The radial and azimuthal indexes of the nanowire are $n_r$ and $n_t$, respectively. The anisotropy parameter is $\eta=n_t/n_r$ and the scattering polar angle is $\theta$. (b) The scattering spectra (normalized scattering cross section $N_{\rm sca}$) with respect to normalized size parameter $\alpha$ for a homogeneous isotropic nanowire ($\eta=1$). Both the total scattering spectrum, and also the partial spectra of solely MD, ED  and EQ  are shown.  The central resonant positions of the first two MDs, EDs and EQs are indicated by $\textbf{M}_{01,02}$ ($\alpha=\beta_{01,02}=0.67,~1.661$), $\textbf{D}_{11,12}$ ($\alpha=\beta_{11,12}=1.02,~1.994$), and $\textbf{Q}_{21,22}$ ($\alpha=\beta_{21,22}=1.384,~2.326$), respectively. The near-field distributions at those points are shown in (d)-(i), where the color-plots correspond to $\textbf{H}_z$, vector-plots correspond to electric fields on the $x-y$ plane, and the dashed circles denote the particle boundaries. (c) The scattering spectra of MDs, EDs, and EQs with anisotropy parameters of $\eta=1.3$ and $\eta=0.7$. For (b) and (c) the azimuthal index is fixed at $n_t=3.5$, as is also the case in Figs.~\ref{fig2}-\ref{fig4}.} \label{fig1}
\end{figure}

\section{Normal scattering of plane waves by radially anisotropic nanowires}

The scattering configuration we study is shown schematically in Fig.~\ref{fig1}(a): a plane wave is normally incident on a homogeneous nanowire (of radius $R$) with wave-vector $\textbf{k}$ along $x$. Considering that the radial anisotropy of refractive index can only affect the TM modes (with magnetic field along nanowire axis direction $z$)~\cite{LIU_Phys.Rev.B_qfactor_2016,LIU_Opt.Lett._qfactor_2016}, here in this study we fix the electric field of the incident wave on plane along $y$.  The refractive indexes of the nanowire are $n_r$ and $n_t$ along radial and azimuthal directions, respectively. The anisotropy parameter is defined as $\eta=n_t/n_r$. For such a case the scattering properties can be analytically calculated, with the normalized scattering cross section (normalized by the single channel scattering limit $2\lambda/\pi$~\cite{Ruan2010_PRL}, where $\lambda$ is the wavelength in the background media) expressed as~\cite{Liu2013_OL2621,chen2012_PRA_anomalous,LIU_Phys.Rev.B_qfactor_2016,LIU_Opt.Lett._qfactor_2016}:
%--------------------------------------------------------------
\begin{equation}
\label{Q_ext}
%Q_{\rm sca} = {2\over {\alpha}}[|a_0|^2 + 2\sum\nolimits_{m = 1}^\infty|a_m|^2],
N_{\rm sca} = \sum\nolimits_{m = -\infty}^\infty|a_m|^2.
\end{equation}
%-------------------------------------------------------------
Here $a_m$ is the scattering coefficient ($m$ is the mode order that characterizes the field distributions along the azimuthal direction), and for TM incident waves: $a_0$ corresponds to the magnetic dipole (MD, $\textbf{M}$); $a_{\pm 1}$ corresponds to the electric dipole (ED, $\textbf{D}$); $a_{\pm 2}$ corresponds to the electric quadrupole(EQ, $\textbf{Q}$) and so on and so forth. It is clear that except the MD with only one scattering channel, all the other electric resonances correspond to two scattering channels which are degenerate as required by the symmetry of the nanowire~\cite{Liu2013_OL2621,chen2012_PRA_anomalous,LIU_Phys.Rev.B_qfactor_2016,LIU_Opt.Lett._qfactor_2016}:
\begin{equation}
\label{scattering_coefficients}
a_{m}=a_{-m}={{n_t\textbf{J} _{\tilde m} (n_t\alpha )\textbf{J}'_{m} (\alpha ) - \textbf{J}_{m} (\alpha )\textbf{J}'_{\tilde m} (n_t\alpha )} \over {n_t\textbf{J} _{\tilde m} (n_t\alpha )\textbf{H} '_{m} (\alpha ) - \textbf{H} _{m} (\alpha )\textbf{J} '_{\tilde m} (n_t\alpha )}},%(m\geq0),
\end{equation}
where $\textbf{J}$ and $\textbf{H}$ are respectively the first-kind Bessel and Hankel functions~\cite{Kerker1969_book}; $\alpha$ is the normalized size parameter $\alpha=kR$; $\tilde m$ is the radial anisotropy-revised function order $\tilde m = m\eta$; and the accompanying primes indicate their differentiation with respect to the entire argument.  According to Eq.~(\ref{scattering_coefficients}), $a_0$ is independent of $\eta$. This is because for the MD, there are only electric fields along the azimuthal direction, and as a result the MD is not affected by the radial anisotropy if the azimuthal index is fixed [see also Figs.~\ref{fig1}(d) and (e)]. By definition, the spectral center of each resonance is located at $\alpha=\beta_{mq}$, which satisfies:
\begin{equation}
\label{resonances}
a_m(\alpha=\beta_{mq})=1.
\end{equation}
Here $q$ is the radial mode number, which corresponds to the number of the field maximum points along the radius[see also Figs.~\ref{fig1}(d)-(i)].

To further exemplify what has been discussed above, in Fig.~\ref{fig1}(b) we show the scattering spectra ($\alpha$ dependence of normalized scattering cross section) of a homogeneous isotropic ($\eta=1$) nanowire. The azimuthal index of the nanowire investigated is fixed at $n_t=3.5$ throughout this work unless otherwise specified. The spectral centers of the first two MDs, EDs, and EQs are indicated respectively by $\textbf{M}_{01,02}$, $\textbf{D}_{11,12}$, and $\textbf{Q}_{21,22}$. The corresponding near-field distributions of those resonances at the points indicated are shown in Figs.~\ref{fig1}(d)-(i), where both the magnetic fields along $z$ ($\textbf{H}_z$, color-plots) and on-plane electric fields (vector-plots) are shown. We emphasize that, to show specifically the characteristic field distributions of each resonance,  in Figs.~\ref{fig1}(d)-(i) we plot the fields associated with each individual resonance only, where the contributions of other resonances are neglected [\textit{e.g.} at $\textbf{Q}_{22}$ the fields associated with the ED and MD are not plotted in Fig.~\ref{fig1}(i)].

The influence of the radial anisotropy on the spectral resonance positions is shown in Fig.~\ref{fig1}(c), where the results for  $\eta=1.3$ and $\eta=0.7$ are summarized. It is clear that with azimuthal index fixed, larger (smaller) $\eta$ will blue-shift (red-shift) the EDs and EQs, with the MD unaffected. It is natural to expect from Fig.~\ref{fig1}(c) that the radial anisotropy can be employed to flexibly overlap resonances of different natures (electric or magnetic), different orders, and/or different radial mode number $q$. This can result in efficient shaping of the superscattering pattern achieved through the resonant interferences of the multipoles co-excited.

\section{Superscattering pattern shaping for radially anisotropic nanowires}

\subsection{Fundamental mechanism and parity of multipolar scattering}

With all the scattering coefficients obtained [see Eq.~(\ref{scattering_coefficients})], we can directly calculate the angular scattering amplitude $\Gamma(\theta)$ as follows~\cite{Liu2013_OL2621,Kerker1969_book}:
%--------------------------------------------------------------
\begin{equation}
\label{scattering_amplitude}
\Gamma(\theta ) = \sqrt {{2 \mathord{\left/{\vphantom {2 {\pi k}}} \right.\kern-\nulldelimiterspace} {\pi k}}} \left|a_0 + 2\sum\nolimits_{m= 1}^\infty  {a_m\cos (m\theta )} \right|,
\end{equation}
%-------------------------------------------------------------
where $k$ is the angular wavenumber(amplitude of wave-vector $\textbf{k}$) in the background medium and $\theta$ is the scattering polar angle with respect to $\textbf{k}$ on the $x-y$ plane [see Fig.~\ref{fig1}(a)]. For the two scattering directions ($\theta$ and $180^{\circ}-\theta$) that are symmetric with respect to $y$ direction and thus in opposite sides (forward or backward) of the scattering circle: considering that $\cos(m\theta)=(-1)^m\cos[m(180^{\circ}-\theta)]$, the scattering amplitudes show odd (even) parity for odd (even)-order resonances. This means that when resonances of different orders are co-excited, there are different types of interferences (constructive or destructive) for scattering amplitudes along different directions, which brings the opportunities for efficient superscattering pattern shaping.

\subsection{Superscattering pattern shaping through resonantly overlapping MDs and EDs}
\begin{figure}
\centerline{\includegraphics[width=8cm]{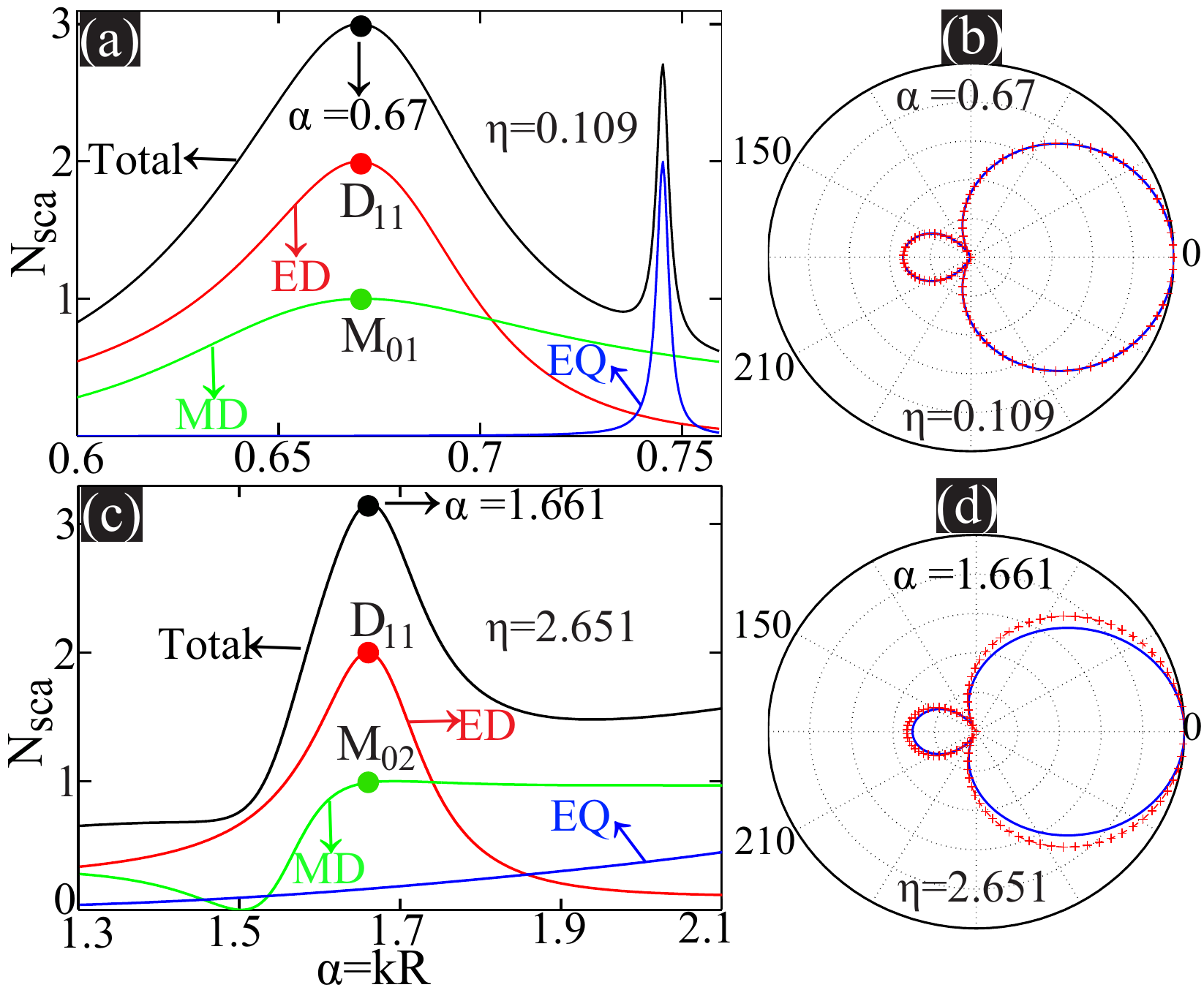}} \caption{Scattering spectra with $\eta=0.109$ (a) and $\eta=2.651$ (c). Both the total spectra (black) and the partial contributions from MDs (green), EDs (red) and EQs (blue) are shown (as is the case for Figs.~\ref{fig3} and \ref{fig4}). The corresponding scattering amplitudes at the overlapping superscattering points indicated are shown respectively by solid blue curves in (b) at $\alpha=0.67$ with $\eta=0.109$ and (d) at $\alpha=1.661$ with $\eta=2.651$. Here we also show the ideal scattering patterns with solely overlapped EDs and MDs by red crosses in (b) and (d).}
\label{fig2}
\end{figure}
For cylindrical scattering particles, the simultaneous superscattering and efficient scattering pattern shaping was firstly achieved in metal-dielectric core-shell nanowires through resonantly overlapping MDs and EDs~\cite{mirzaei2013cloaking,Liu2013_OL2621}. Since for 2D nanowires, the number of scattering channels of MD is only half of the ED scattering channels [Eq.~(\ref{scattering_amplitude})], the scattering is suppressed at other angles ($\theta  = {120^{\circ}}$ and ${240^{\circ}}$)~\cite{Liu2013_OL2621} rather than at the backward direction ($\theta  = {180^{\circ}}$) for 3D metal-dielectric core-shell spherical particles~\cite{Liu2012_ACSNANO}. Here we show the resonant overlapping of EDs and MDs within a homogenous radially anisotropic nanowire, and the results are summarized in  Fig.~\ref{fig2}.  Figure~\ref{fig2}(a) shows the scattering spectra (both total and partial contributions) for $\eta=0.109$, where it is clear that the anisotropy induced red shift of the first ED ($\textbf{D}_{11}$) enables its overlapping with the spectrally-fixed first MD ($\textbf{M}_{01}$), leading to effective superscattering beyond the single channel scattering limit ($N_{\rm sca}=1$)~\cite{Ruan2010_PRL}. The scattering amplitude at the overlapping superscattering point ($\alpha=0.67$) is shown in Fig.~\ref{fig2}(b) by solid curve. Similar to what has been achieved for core-shell nanowires~\cite{Liu2013_OL2621}, the scattering is suppressed at the backward half scattering circle while enhanced in the forward half scattering circle. This is due to the fact that: (i) the scatterings of ED and MD exhibit different parities (odd and even respectively) and (ii) at the forward direction the two resonantly overlapped multipoles are always in phase according to the optical theorem~\cite{Kerker1969_book}.  For comparison, we also show in Fig.~\ref{fig2}(b) by crosses the ideal scattering pattern of resonant overlapping of ED and MD only with all other multipoles neglected ($a_0=a_{\pm 1}=1, a_{|m|>1}=0$): $\Gamma(\theta)\propto|1+\cos(\theta)|$.  It is clear in Fig.~\ref{fig2}(b) that the results of the two scenarios agree perfectly well, as the other multipoles can be effectively neglected at $\alpha=0.67$ for $\eta=0.109$ [see Fig.~\ref{fig2}(a)].

The ED can be also engineered to resonantly overlap with the MD with larger anisotropy parameter $\eta=2.651$, as is shown in Fig.~\ref{fig2}(c), where the first ED ($\textbf{D}_{11}$) coincides spectrally with the second MD ($\textbf{M}_{02}$) at $\alpha=1.661$. The corresponding scattering amplitude is shown in Fig.~\ref{fig2}(d) (solid curve), which is slightly different from the ideal case (crosses). This is due the fact that at $\alpha=1.661$, the contributions of EQ are noneligible [see Fig.~\ref{fig2}(c)].

\subsection{Superscattering pattern shaping through resonantly overlapping MDs and EQs}
\begin{figure}
\centerline{\includegraphics[width=8cm]{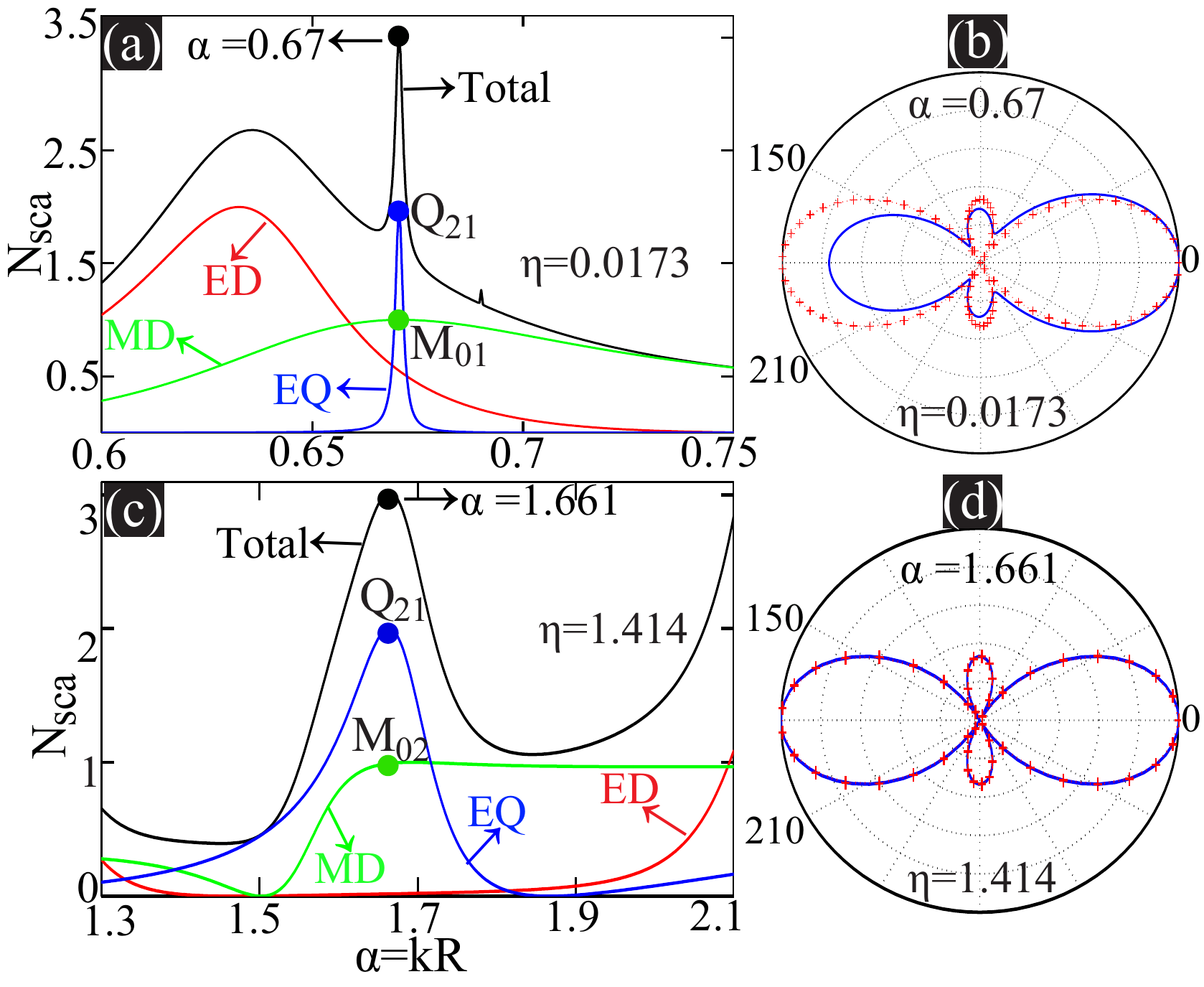}} \caption{Scattering spectra with $\eta=0.0173$ (a) and $\eta=1.414$ (c). The corresponding scattering amplitudes at the overlapping superscattering points indicated are shown respectively by solid blue curves in (b) at $\alpha=0.67$ with $\eta=0.0173$  and (d) of $\alpha=1.661$ with $\eta=1.414$. Here we also show the ideal scattering patterns with solely overlapped EQs and MDs by red crosses in (b) and (d).}
\label{fig3}
\end{figure}
According to Fig.~\ref{fig1}(c), similar to EDs, the spectral positions of EQs are also sensitive to $\eta$ and thus the radial anisotropy can be employed to resonantly overlap EQs with MDs too, resulting in also effective superscattering. This is shown in Fig.~\ref{fig3}(a) with $\eta=0.0173$ and Fig.~\ref{fig3}(c) with $\eta=1.414$. For the former case the first EQ ($\textbf{Q}_{21}$) is tuned to overlap with first MD ($\textbf{M}_{01}$), while for the latter case it is the first EQ overlapped with the second MD ($\textbf{M}_{02}$). The corresponding scattering amplitudes at the overlapping superscattering points are shown by solid curves in Fig.~\ref{fig3}(b) ($\alpha=0.67$ with with $\eta=0.0173$) and (d) ($\alpha=1.661$ with $\eta=1.414$) together with the ideal scattering patterns (crosses) with only resonantly overlapped EQ and MD: $\Gamma(\theta)\propto|1+\cos(2\theta)|$. The discrepancy in Fig.~\ref{fig3}(b) originates from the noneligible contributions of ED [see Fig.~\ref{fig3}(a)] while in the Fig.~\ref{fig3}(d) the two sets of results are perfectly matched as there are no other multipolar contributions at the overlapping superscattering point [see Fig.~\ref{fig3}(c)]. In contrast to the scattering patterns of overlapped MDs and EDs shown in Figs.~\ref{fig2}(b) and (d), the scattering amplitudes are symmetric in the forward and backward half scattering circles, which is induced by the same even parity of the scatterings of the MD and EQ. The complete destructive interferences still significantly suppress [Fig.~\ref{fig3}(b)] or fully eliminate [Fig.~\ref{fig3}(d)] the scattering at $\theta=60^{\circ},~120^{\circ},~240^{\circ}$, and $300^{\circ}$.

\subsection{Superscattering pattern shaping through resonantly overlapping EDs and EQs}
\begin{figure}
\centerline{\includegraphics[width=8cm]{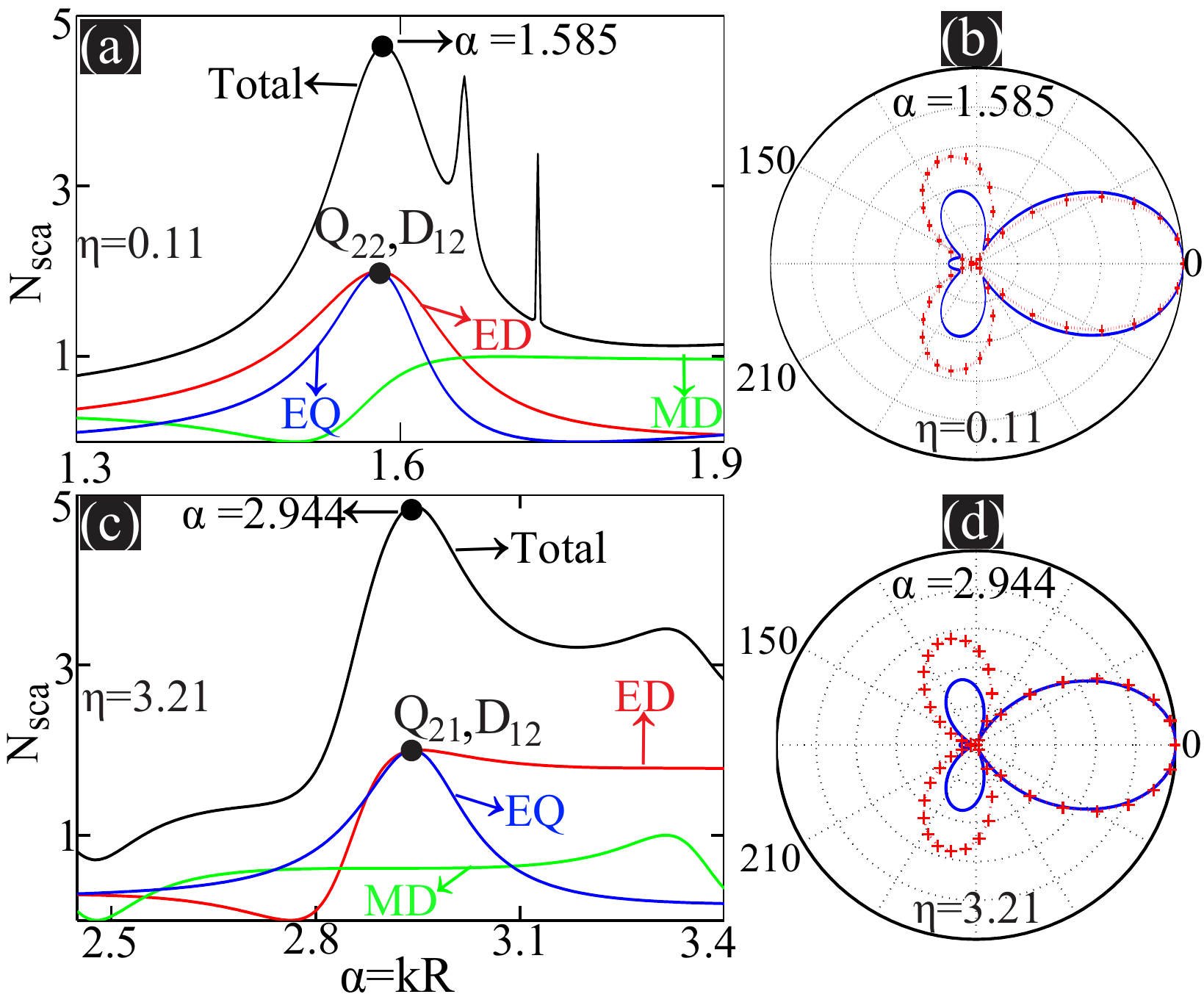}} \caption{Scattering spectra with $\eta=0.11$ (a) and $\eta=3.21$ (c). The corresponding scattering amplitudes at the overlapping superscattering points are shown respectively by solid blue curves in (b) at $\alpha=1.585$ with $\eta=0.11$  and (d) of $\alpha=2.944$ with $\eta=3.21$. Here we also show the ideal scattering patterns with pure overlapped EQs and EDs by red crosses in (b) and (d).}
\label{fig4}
\end{figure}
Up to now, we have managed to resonantly overlap spectrally $\eta$-sensitive electric multipoles with spectrally fixed MDs relying on radial anisotropy. Though both EDs and EQs will red-shift (blue-shift) with smaller (larger) $\eta$ [see Fig.~\ref{fig1}(c)], it is also possible to overlap them as has been shown in Fig.~\ref{fig4}(a) [$\eta=0.11$, second EQ ($\textbf{Q}_{22}$) with second ED ($\textbf{D}_{12}$)], and in Fig.~\ref{fig4}(c) [$\eta=3.21$, first EQ ($\textbf{Q}_{21}$) with second ED]. The scattering patterns at the overlapping superscattering points ($\alpha=1.585,~2.944$) are shown in Figs.~\ref{fig4}(b) and (d) respectively by solid curves, where the ideal scattering amplitudes with pure resonantly overlapped ED and EQ [$\Gamma(\theta)\propto|\cos(\theta)+\cos(2\theta)|$] are also shown by crosses. The discrepancies exist for both scenarios, as at both overlapping superscattering points the contributions from MDs cannot be neglected [see Figs.~\ref{fig4}(b) and (d)]. Similar to what have been shown in Figs.~\ref{fig2}(b) and (d), the symmetry of the scattering amplitudes in the forward and backward half scattering circles are also broken since the parities of the ED and EQ scattering are different (odd and even respectively). It is clear that the interferences of ED and EQ can significantly suppress the backward reflection and also scattering at other two angles in the forward half scattering circle ($\theta=60^{\circ}$ and $300^{\circ}$).

\begin{figure*}
\centerline{\includegraphics[width=12cm]{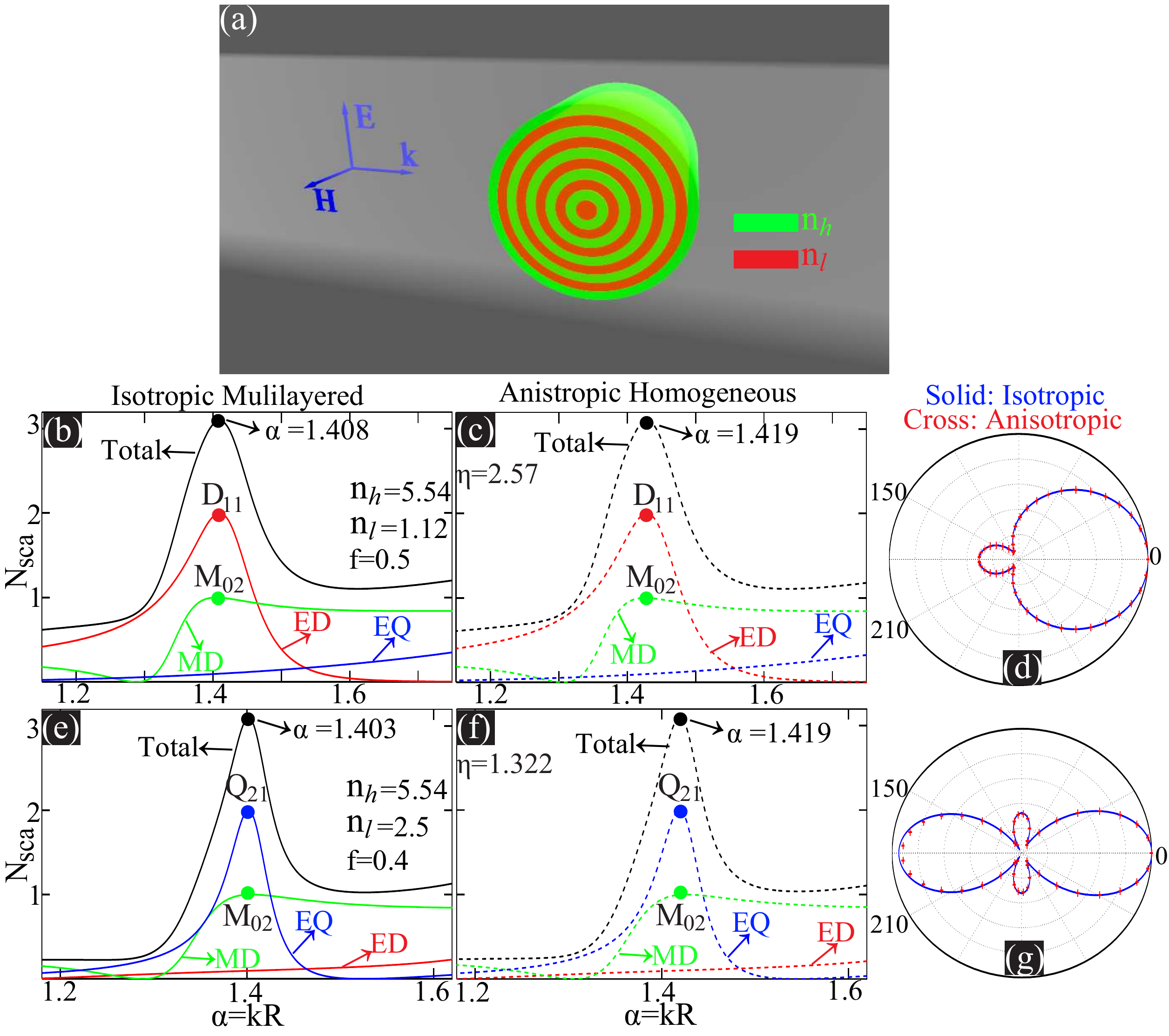}} \caption{(a) Schematic of a multi-layered nanowire made of alternate isotropic layeres of refractive indexes $n_1$ (each layer width $d_1$) and $n_2$ (each layer width $d_2$), and thus $f=d_1/(d_1+d_2)$. The radius of the whole nanowire is $R_m$. Scattering spectra of $30$-layered nanowire are shown in (b) with $n_1=5.54$, $n_2=1.12$, $d_1=d_2=R_m/30$, $f=0.5$, and in (c) with $n_1=5.54$, $n_2=2.5$, $d_1=2d_2/3=2R_m/75$, $f=0.4$. The scattering spectra of the corresponding anisotropic homogeneous nanowires are shown respectively in (c) with $n_t=4$  and $\eta=2.57$, and in (f)  with $n_t=4$ and $\eta=1.322$.  The first ED ($\textbf{D}_{11}$) and the second MD ($\textbf{M}_{02}$) are resonantly overlapped in (b) at $\alpha=1.408$ and in (c) at $\alpha=1.419$. The scattering amplitudes at those superscattering points are shown in (d); The first EQ ($\textbf{Q}_{21}$) and the second MD are resonantly overlapped in (e) at $\alpha=1.403$ and in (f) at $\alpha=1.419$. The scattering amplitudes at those superscattering points are shown in (g).}
\label{fig5}
\end{figure*}

\section{Superscattering pattern shaping for multi-layered isotropic nanowires with effective radial anisotropy}
In the discussions above about homogeneous anisotropic nanowires, we have employed unusual anisotropy parameters that are inaccessible for natural materials.  Nevertheless, the recent rapid development of the field of metamaterials has provided lots of opportunities to obtain extreme anisotropy parameters in various artificial structures~\cite{choi2011_nature_terahertz,poddubny2013_NP_hyperbolic}.  For example, effectively large radial anisotropy parameters can be obtained in a multi-layered nanowire that is shown schematically in Fig.~\ref{fig5}(a)\cite{jahani_alldielectric_2016,LIU_Phys.Rev.B_qfactor_2016,LIU_Opt.Lett._qfactor_2016,poddubny2013_NP_hyperbolic}. Such a core-shell structure is made of alternating isotropic layers of two refractive indexes of $n_1$ and $n_2$. According to the effective medium theory, the whole multi-layered isotropic structure can be viewed effectively as a homogeneous radially anisotropic nanowire with radial and azimuthal indexes expressed respectively as~\cite{LIU_Phys.Rev.B_qfactor_2016,LIU_Opt.Lett._qfactor_2016,LIU_Sci.Rep._unidirectional_2016}:
\begin{equation}
\label{effective1}
\begin{split}
&n_{\rm {r}} =n_1n_2/\sqrt {(1 - f)n_1^2  + fn_2^2 },\\
&n_{\rm {t}} =\sqrt {fn_l^2  + (1 - f)n_2^2},
\end{split}
\end{equation}
where $f$ is the filling factor of the layer of index $n_1$ (the overall thickness of all the layers of index $n_1$ divided by the radius of the whole structure $R_m$).

Firstly we study a $30$-layered nanowire made of alternating dielectric layers of $n_1=5.54$ (each layer width is $d_1=R_m/30$) and $n_2=1.12$ (each layer width is $d_2=R_m/30$),  and thus $f=d_1/(d_1+d_2)=0.5$.  According to Eq.~(\ref{effective1}), the corresponding effective parameter is $\eta=2.57$ and $n_t=4$.  The scattering spectra of such multi-layered nanowire (which can be analytically calculated based on generalized Mie theory~\cite{Kerker1969_book}) and of its corresponding homogeneous anisotropic nanowire are shown in  Figs.~\ref{fig5}(b) and (c), respectively. The two sets of spectra agree quite well (validating the effectiveness of effective medium theory employed) and for both cases the first ED ($\textbf{D}_{11}$) can be tuned to resonantly overlap with the second MD ($\textbf{M}_{02}$), at $\alpha=1.408$ and $\alpha=1.419$ respectively. The spectra of the anisotropic nanowire [Fig.~\ref{fig5}(c)] is blue-shifted compared to those of the isotropic one [Fig.~\ref{fig5}(b)], and such discrepancies can be fully eliminated by decreasing each consisting isotropic layer width, as is also the case for Figs.~\ref{fig5}(e) and (f). The scattering amplitudes at the overlapping superscattering points are shown in  Figs.~\ref{fig5}(d) for both cases, which are basically the same as what is shown in Fig.~\ref{fig2}(d).

We also study another similar $30$-layered nanowire with $n_1=5.54$ ($d_1=2R_m/75$), $n_2=2.5$ ($d_2=3R_m/75$) and $f=0.4$, which corresponds to effective parameters of $\eta=1.322$ and $n_t=4$. The scattering spectra for both isotropic and anisotropic nanowires are shown in Figs.~\ref{fig5}(e) and (f), where it is clear that the first EQ ($\textbf{Q}_{21}$) and the second MD ($\textbf{M}_{02}$) are resonantly overlapped. The scattering amplitudes at the overlapping superscattering points (isotropic nanowire: $\alpha=1.403$ ; anisotropic nanowire: $\alpha=1.419$) are shown in Fig.~\ref{fig5}(g), which are more or less that same as that shown in Fig.~\ref{fig3}(d).

It is worth mentioning that here we have confined our studies to all-dielectric nanowires with $n_1>0$ and $n_2>0$, which according to Eq.~(\ref{effective1}) leads to $\eta=n_t/n_r>1$. Nevertheless, when we go beyond the all-dielectric regime, such as incorporating metals into the multi-layered configurations, $\eta<1$ and even more exotic anisotropic parameters can be obtained~\cite{LIU_Sci.Rep._unidirectional_2016,poddubny2013_NP_hyperbolic}. As a result, it is expected that other superscattering features shown in Figs.~\ref{fig2}-\ref{fig4} can be also observed within multi-layered isotropic nanowires.

\section{Conclusions and Outlook}
In conclusion, we investigate the scattering properties of radially anisotropic nanowires and  have achieved superscattering with engineered angular distributions. The efficient shaping of the superscattering pattern achieved originates from the resonant overlapping and different sorts of interferences between the electric and magnetic multipoles co-excited, which is made possible by incorporating radial anisotropy of refractive index into homogeneous nanowires.  We further demonstrate that the large anisotropy parameters required that are inaccessible for natural materials can be realized in artificial multi-layered isotropic core-shell nanowires, where multipolar interference-induced superscattering manipulation can be also obtained.

Here in this work, we confine our investigations to lower order resonances up to quadrupoles ($m\leq 2$) with small radial mode numbers ($q\leq 2$), and have discussed only the case of two overlapped resonances. Similar studies can certainly be extended to higher order modes with larger radial mode numbers and to the cases of more than two overlapped resonances~\cite{Liu2014_ultradirectional}, where we expect extra flexibilities for more efficient superscattering pattern shaping. Moreover, the principle we have revealed here can be also applied to particle clusters and periodic arrays~\cite{Lukyanchuk2011_NM,Miroshnichenko2010_RMP,Liu2012_PRB}, where the eigenmodes of the whole system can be tuned by the index anisotropy and this may render much more opportunities for superscattering pattern shaping replying on the extra dimension of freedom of inter-particle interaction control.  At the same time, other kinds of anisotropy such as magnetic anisotropy can be also employed~\cite{KRUK_NatCommun_magnetic_2016}. We anticipate that the approach based on anisotropy to achieve simultaneous superscattering and efficient scattering angular distribution control can not only play a significant role in various applications replying on resonant particle scattering, but also bring new stimuli for the emerging fields of topological photonics, and low-dimensional photonics which involves the novel two-dimensional and topological materials that show intrinsically huge anisotropy.

%\section*{Acknowledgments}
We are indebted to A. E. Miroshnichenko and Y. S. Kivshar for many useful discussions and suggestions, and acknowledge the financial support from the National Natural Science Foundation of China (Grant number: $11404403$), and the Outstanding Young Researcher Scheme of National University of Defense Technology.

%\section*{References}
%\bibliographystyle{osajnl}
%\bibliography{References_scattering}

\begin{thebibliography}{10}
\newcommand{\enquote}[1]{``#1''}

\bibitem{Liu2014_CPB}
W.~Liu, A.~E. Miroshnichenko, and Y.~S. Kivshar, \enquote{Control of light
  scattering by nanoparticles with optically-induced magnetic responses,} Chin.
  Phys. B \textbf{23}, 047806 (2014).

\bibitem{jahani_alldielectric_2016}
S.~Jahani and Z.~Jacob, \enquote{All-dielectric metamaterials,} Nat. Nanotech.
  \textbf{11}, 23--26 (2016).

\bibitem{KUZNETSOV_Science_optically_2016}
A.~I. Kuznetsov, A.~E. Miroshnichenko, M.~L. Brongersma, Y.~S. Kivshar, and
  B.~Luk'yanchuk, \enquote{Optically resonant dielectric nanostructures,}
  Science \textbf{354}, 2472 (2016).

\bibitem{SMIRNOVA_Optica_multipolar_2016}
D.~Smirnova and Y.~S. Kivshar, \enquote{Multipolar nonlinear nanophotonics,}
  Optica \textbf{3}, 1241--1255 (2016).

\bibitem{LIU_ArXivPrepr.ArXiv160901099_multipolar_2016}
W.~Liu and Y.~S. Kivshar, \enquote{Multipolar interference effects in
  nanophotonics,} Phil. Trans. R. Soc. A \textbf{375}, 20160317 (2017).

\bibitem{Ruan2010_PRL}
Z.~C. Ruan and S.~H. Fan, \enquote{Superscattering of light from subwavelength
  nanostructures,} Phys. Rev. Lett. \textbf{105}, 013901 (2010).

\bibitem{Liu2014_arXiv_Geometric}
W.~Liu, R.~F. Oulton, and Y.~S. Kivshar, \enquote{Geometric interpretations for
  resonances of plasmonic nanoparticles,} Sci. Rep. \textbf{5}, 12148 (2015).

\bibitem{LI_IEEEJ.Sel.Top.QuantumElectron._design_2017}
R.~Li, B.~Zheng, X.~Lin, R.~Hao, S.~Lin, W.~Yin, E.~Li, and H.~Chen,
  \enquote{Design of {{Ultracompact Graphene}}-{{Based Superscatterers}},} IEEE
  J. Sel. Top. Quantum Electron. \textbf{23}, 1--8 (2017).

\bibitem{Liu2012_ACSNANO}
W.~Liu, A.~E. Miroshnichenko, D.~N. Neshev, and Y.~S. Kivshar,
  \enquote{Broadband unidirectional scattering by magneto-electric core-shell
  nanoparticles,} ACS Nano \textbf{6}, 5489--5497 (2012).

\bibitem{mirzaei2013cloaking}
A.~Mirzaei, I.~V. Shadrivov, A.~E. Miroshnichenko, and Y.~S. Kivshar,
  \enquote{Cloaking and enhanced scattering of core-shell plasmonic nanowires,}
  Opt. Express \textbf{21}, 10454 (2013).

\bibitem{Liu2013_OL2621}
W.~Liu, A.~E. Miroshnichenko, R.~F. Oulton, D.~N. Neshev, O.~Hess, and Y.~S.
  Kivshar, \enquote{Scattering of core-shell nanowires with the interference of
  electric and magnetic resonances,} Opt. Lett. \textbf{38}, 2621--2624 (2013).

\bibitem{Liu2014_ultradirectional}
W.~Liu, J.~Zhang, B.~Lei, H.~Ma, W.~Xie, and H.~Hu, \enquote{Ultra-directional
  forward scattering by individual core-shell nanoparticles,} Opt. Express
  \textbf{22}, 16178 (2014).

\bibitem{Staude2013_acsnano}
I.~Staude, A.~E. Miroshnichenko, M.~Decker, N.~T. Fofang, S.~Liu, E.~Gonzales,
  J.~Dominguez, T.~S. Luk, D.~N. Neshev, I.~Brener, and Y.~Kivshar,
  \enquote{Tailoring directional scattering through magnetic and electric
  resonances in subwavelength silicon nanodisks,} ACS Nano \textbf{7}, 7824
  (2013).

\bibitem{LUKYANCHUK_ACSPhotonics_optimum_2015}
B.~S. Luk'yanchuk, N.~V. Voshchinnikov, R.~Paniagua-Dom{\'\i}nguez, and A.~I.
  Kuznetsov, \enquote{Optimum {{Forward Light Scattering}} by {{Spherical}} and
  {{Spheroidal Dielectric Nanoparticles}} with {{High Refractive Index}},} ACS
  Photonics \textbf{2}, 993--999 (2015).

\bibitem{Liu2015_OE_Ultra}
W.~Liu, \enquote{Ultra-directional super-scattering of homogenous spherical
  particles with radial anisotropy,} Opt. Express \textbf{23}, 14734--14743
  (2015).

\bibitem{LIU_Sci.Rep._unidirectional_2016}
W.~Liu, B.~Lei, J.~Shi, and H.~Hu, \enquote{Unidirectional superscattering by
  multilayered cavities of effective radial anisotropy,} Sci. Rep. \textbf{6},
  34775 (2016).

\bibitem{Kerker1969_book}
M.~Kerker, \emph{The scattering of light, and other electromagnetic radiation}
  (Academic Press, New York, 1969).

\bibitem{Novotny2012_book}
L.~Novotny and B.~Hecht, \emph{Principles of nano-optics} (Cambridge University
  Press, Cambridge, 2012), 2nd ed.

\bibitem{ZHAI_Science_scalablemanufactured_2017-1}
Y.~Zhai, Y.~Ma, S.~N. David, D.~Zhao, R.~Lou, G.~Tan, R.~Yang, and X.~Yin,
  \enquote{Scalable-manufactured randomized glass-polymer hybrid metamaterial
  for daytime radiative cooling,} Science \textbf{355}, 1062--1066 (2017).

\bibitem{LIU_Phys.Rev.B_qfactor_2016}
W.~Liu, A.~E. Miroshnichenko, and Y.~S. Kivshar, \enquote{Q-factor enhancement
  in all-dielectric anisotropic nanoresonators,} Phys. Rev. B \textbf{94},
  195436 (2016).

\bibitem{LIU_Opt.Lett._qfactor_2016}
W.~Liu, B.~Lei, and A.~E. Miroshnichenko, \enquote{Q-factor and absorption
  enhancement for plasmonic anisotropic nanoparticles,} Opt. Lett. \textbf{41},
  3563 (2016).

\bibitem{chen2012_PRA_anomalous}
H.~Chen and L.~Gao, \enquote{Anomalous electromagnetic scattering from radially
  anisotropic nanowires,} Phys. Rev. A \textbf{86}, 033825 (2012).

\bibitem{choi2011_nature_terahertz}
M.~Choi, S.~H. Lee, Y.~Kim, S.~B. Kang, J.~Shin, M.~H. Kwak, K.~Y. Kang, Y.~H.
  Lee, N.~Park, and B.~Min, \enquote{A terahertz metamaterial with unnaturally
  high refractive index,} Nature \textbf{470}, 369--373 (2011).

\bibitem{poddubny2013_NP_hyperbolic}
A.~Poddubny, I.~Iorsh, P.~Belov, and Y.~Kivshar, \enquote{Hyperbolic
  metamaterials,} Nat. Photon. \textbf{7}, 948--957 (2013).

\bibitem{Lukyanchuk2011_NM}
B.~Luk'yanchuk, N.~I. Zheludev, S.~A. Maier, N.~J. Halas, P.~Nordlander,
  H.~Giessen, and C.~T. Chong, \enquote{The fano resonance in plasmonic
  nanostructures and metamaterials,} Nat. Mater. \textbf{9}, 707 (2010).

\bibitem{Miroshnichenko2010_RMP}
A.~E. Miroshnichenko, S.~Flach, and Y.~S. Kivshar, \enquote{Fano resonances in
  nanoscale structures,} Rev. Mod. Phys. \textbf{82}, 2257 (2010).

\bibitem{Liu2012_PRB}
W.~Liu, A.~E. Miroshnichenko, D.~N. Neshev, and Y.~S. Kivshar,
  \enquote{Polarization-independent fano resonances in arrays of core-shell
  nanoparticles,} Phys. Rev. B \textbf{86}, 081407(\textbf{R}) (2012).

\bibitem{KRUK_NatCommun_magnetic_2016}
S.~S. Kruk, Z.~J. Wong, E.~Pshenay-Severin, K.~O'Brien, D.~N. Neshev, Y.~S.
  Kivshar, and X.~Zhang, \enquote{Magnetic hyperbolic optical metamaterials,}
  Nat. Commun. \textbf{7}, 11329 (2016).

\end{thebibliography}

%==========================
\end{document}